\documentstyle[aps,twocolumn,epsf]{revtex}
\begin{document}
\title{Entanglement of photons}
\author{S.J. van Enk\\
Bell Labs, Lucent Technologies,
Room 2C-401\\
600-700 Mountain Ave,
Murray Hill NJ 07974}
\maketitle
\begin{abstract}
It is argued that the title of this paper represents a misconception.
Contrary to widespread beliefs it is electromagnetic field modes that are ``systems'' and can be entangled, not photons. 
The amount of entanglement in a given state is shown to depend on redefinitions of the modes; we calculate the minimum and maximum over all such redefinitions for several examples. 
\end{abstract}
\medskip
\section{Introduction}
If we have two quantum systems, then we all know that their joint quantum state may be entangled. This is the case when there is no way to write that state as a product state (or a mixture of product states) of the two systems. The systems may be particles such as electrons, atoms, or ions. In the quantum theory of light, which is a second-quantized theory, it is not the ``particles'', the photons, that form systems, but rather the EM field modes. 
A (pure) quantum state of a mode is an arbitrary superposition 
\begin{equation}\label{Psi}
|\Psi\rangle=\sum_{n=0}^{\infty} \alpha_n |n\rangle
\end{equation}
with the integer $n$ the number of photons in that mode and $\alpha_n$ arbitrary complex numbers only restricted by the normalization condition.

Two field modes $A$ and $B$ are described by spatially dependent (vector) mode functions $\vec{F}_{j}(\vec{r})$ and by creation and annihilation operators $a^{\dagger}_j,a_j$ for $j=A,B$ \cite{E}.
These field modes are independent systems if the mode functions are orthogonal, 
\begin{equation}\label{o1}
\int {\rm d}\vec{r} \vec{F}^*_i(\vec{r})\cdot\vec{F}_j(\vec{r})=\delta_{ij}.
\end{equation}
The mode operators then satisfy the standard bosonic commutation relations
\begin{equation}\label{o2}
[a_i,a^{\dagger}_j]=\delta_{ij},
\end{equation}
and the number states $|n\rangle_j$ of mode $j$ are
 given by 
\begin{equation}
|n\rangle_j=\frac{(a^{\dagger}_j)^n}{\sqrt{n!}}|0\rangle_j.
\end{equation}
Before moving on to more interesting states, we first consider a simple example that shows that entanglement is between modes but not between photons. If we have a state
\begin{equation}\label{s1}
|\psi\rangle=(|0\rangle_A|1\rangle_B
+|1\rangle_A|0\rangle_B)/\sqrt{2},
\end{equation}
there is one ebit of entanglement\cite{ent} between the modes $A$ and $B$, although there is only one photon. In particular, when modes $A$ and $B$ are located in different spatial regions
then the entanglement between $A$ and $B$ is truly nonlocal, may be used for teleportation, and will violate Bell inequalities.

On the other hand, the same example shows that the amount of entanglement depends on the definition of the modes. Namely, given a set of mode functions $\vec{F}_j$ and corresponding mode operators $a^{\dagger}_j$  we are free to define new mode operators and functions $j'$ by \cite{loudon}
\begin{eqnarray}\label{redef}
a^{\dagger}_{j'}&=&\sum_j U_{j'j}a^{\dagger}_j,\nonumber\\
\vec{F}_{j'}&=&\sum_j U_{j'j}\vec{F}_j,
\end{eqnarray}
where $U_{j'j}$ is an arbitrary unitary matrix, so that conditions (\ref{o1}) and (\ref{o2}) are still satisfied \cite{E2}.
In particular, we may rewrite the state (\ref{s1}) as
\begin{equation}\label{s2}
|\psi\rangle=|1\rangle_{A'}|0\rangle_{B'}
\end{equation}
in terms of new modes $A',B'$, where
\begin{eqnarray}\label{bishop}
a^{\dagger}_{A'}&=&(a^{\dagger}_{A}+a^{\dagger}_{B})/\sqrt{2},\nonumber\\
a^{\dagger}_{B'}&=&(a^{\dagger}_{A}-a^{\dagger}_{B})/\sqrt{2}.
\end{eqnarray}
There is no entanglement in the state (\ref{s2}) between systems $A'$ and $B'$. For example, if $A$ and $B$ correspond to horizontal and vertical polarization (with all other quantum numbers being the same), then the state (\ref{s1}) describes a single diagonally polarized photon which one normally would not call entangled.

It is important to note that, depending on the context, not all redefinitions of the form (\ref{redef}) may be relevant. Entanglement between two particular modes and the very definitions of the modes are only useful if one can perform measurements on those modes.  
With this criterion, two spatially separated modes $A$ and $B$ cannot be redefined in a useful way if only local measurements are possible.
If we have two sets of modes $\{A_1\ldots A_n\}$ and $\{B_1\ldots B_m\}$ located in two different spatial regions
then within each set redefinitions are allowed. However, such ``local'' redefinitions, as expected, cannot change the nonlocal entanglement between the two sets of modes.
In particular, in most (if not all) experiments entanglement is created between different spatial modes. For example, 
in downconversion experiments \cite{downc} the so-called signal and idler modes propagate in different directions and may be entangled.
Similarly, a two-mode squeezed state \cite{twomode} 
is produced by splitting two squeezed light beams on a beam splitter \cite{BS} with the two output beams propagating in different directions. In such cases there is a preferred set of modes, and for states with a definite number of photons one does get the correct amount of entanglement between those modes even if one treats photons as systems.

Perhaps more importantly, one can view the redefinition (\ref{redef}) as a unitary operation that one can apply actively in order to create (or destroy) entanglement. For example, a polarizing beam splitter may be used to turn the state (\ref{s2}) into the state (\ref{s1}). In general, if all modes in the problem are spatialy different then all unitary operations of the form (\ref{redef}) can be performed with just linear optics (beam splitters and mirrors). Other cases not covered by polarization or spatial degrees of freedom are discussed in the Examples Section (specifically, in Section \ref{orb}).

We will discuss several examples of pure states and give the maximum and minimum entanglement possible by varying over all possible unitary redefinitions/operations of the form (\ref{redef}). We denote these quantities by $E_{{\rm max}}$ and $E_{{\rm min}}$. 
We only consider transformations that leave the number of modes the same (or at least that do not increase the number). Otherwise, any state with more than a single photon can be transformed into one with an arbitrarily large amount of entanglement by introducing an arbitrarily large number of modes. We also restrict ourselves to {\em bipartite} entanglement.
In the simple example of the state (\ref{s1})  
it is easy to see that under these conditions $E_{{\rm min}}=0$ and $E_{{\rm max}}=1$. 
\section{Examples}
We first consider states where the total photon number is fixed and equal to 2.
\subsection{$|20\rangle+|02\rangle$}\label{pawn}
Consider a state where we have two modes and two photons in a state
\begin{equation}\label{s02}
|\Psi\rangle=(  |2\rangle_{A}|0\rangle_{B}+ |0\rangle_{A}|2\rangle_{B})   /\sqrt{2}.
\end{equation}
For example, the two modes might describe horizontal and vertical polarization (with the remaining quantum numbers the same for both modes).
Writing this state, incorrectly, as
\begin{equation}
|\Psi\rangle=(|\updownarrow\rangle|\updownarrow\rangle+
|\leftrightarrow\rangle|\leftrightarrow\rangle)/\sqrt{2}\,\,\,\,\,\,({\rm wrong!}),
\end{equation}
we would conclude this state has one ebit of entanglement and we might even think that that is the only correct answer (Note here that this notation is usually meant to imply that the two kets refer to different spatial modes; in that case the notation is not wrong, just a little dangerous). However, on the one hand the state (\ref{s02}) can be rewritten as a product state
\begin{equation}\label{s02b}
|\Psi\rangle=|1\rangle_{A'}|1\rangle_{B'},
\end{equation}
so that $E_{{\rm min}}=0$,
by redefining
\begin{eqnarray}
a_{A'}^{\dagger}&=&(a_A^{\dagger}+ia_B^{\dagger})/\sqrt{2},\nonumber\\
a_{B'}^{\dagger}&=&(a_A^{\dagger}-ia_B^{\dagger})/\sqrt{2}.
\end{eqnarray}
Thus, when the original modes $A$ and $B$ describe linear polarization (as above), the state (\ref{s02}) is equivalent to a state containing one left-hand circularly polarized photon and one right-hand circularly polarized photon.
On the other hand, we have $E_{{\rm max}}=\log_2(3)$;
we can also rewrite the state as
\begin{equation}\label{s02c}
|\Psi\rangle=(|1\rangle_{A''}|1\rangle_{B''}+ |2\rangle_{A''}|0\rangle_{B''}+ |0\rangle_{A''} |2\rangle_{B''}                  )/\sqrt{3}.
\end{equation}
The unitary operation accomplishing this transformation is less obvious now, but one of them is
\begin{eqnarray}
a_{A''}^{\dagger}&=&(a_A^{\dagger}+xa_B^{\dagger})/\sqrt{2},\nonumber\\
a_{B''}^{\dagger}&=&(a_A^{\dagger}-xa_B^{\dagger})/\sqrt{2},
\end{eqnarray}
where $x=1/3+2\sqrt{2}i/3$. One cannot get more than $\log_2(3)$ ebits of entanglement between 2 single modes, because the reduced density matrix of any mode can be at most of rank 3.
\subsection{$|0110\rangle+|1001\rangle$}
Now consider two photons in 4 modes $A\ldots D$. Suppose we have a state of the form
\begin{equation}\label{s0110}
|\Psi\rangle=(|0\rangle|1\rangle|1\rangle|0\rangle
+|1\rangle|0\rangle|0\rangle|1\rangle)/\sqrt{2},
\end{equation}
where we left out the bothersome subscripts $ABCD$ to indicate the modes.
By writing this state in dangerous notation as
\begin{equation}\label{s0110b}
|\Psi\rangle=(|B\rangle|C\rangle+|A\rangle|D\rangle)/\/\sqrt{2},
\end{equation}
we may conclude that the entanglement between the photons is 1 ebit. Indeed, when we restrict ourselves to bipartite entanglement between two pairs of modes, it turns out that $E_{{\rm min}}=1$. But in this case $E_{{\rm max}}=2$. Namely, we can rewrite the state (\ref{s0110}) as
\begin{eqnarray}
|\Psi\rangle&=&(|0\rangle|1\rangle|1\rangle|0\rangle
+|1\rangle|0\rangle|0\rangle|1\rangle\nonumber\\
&+&|0\rangle|0\rangle[|0\rangle|2\rangle+|2\rangle|0\rangle]/\sqrt{2}
\nonumber\\
&-&[|0\rangle|2\rangle+|2\rangle|0\rangle]|0\rangle|0\rangle/\sqrt{2}
)/2,
\end{eqnarray}
which manifestly possesses 2 ebits of entanglement between the first pair of modes $A',B'$ and the second pair $C',D'$.
The corresponding transformation is
\begin{eqnarray}
a_{B'}^{\dagger}&=&x_+a_B^{\dagger}-x_-a_C^{\dagger},\nonumber\\
a_{C'}^{\dagger}&=&x_-a_B^{\dagger}+x_+a_C^{\dagger},\nonumber\\
a_{A'}^{\dagger}&=&x_+a_A^{\dagger}-x_-a_D^{\dagger},\nonumber\\
a_{D'}^{\dagger}&=&x_-a_A^{\dagger}+x_+a_D^{\dagger},\nonumber\\
\end{eqnarray}
with $x_{\pm}=\sqrt{1/2\pm\sqrt{2}/4}$. 
Since $\rho_{A'B'}$ can be at most of rank 4, 2 ebits of entanglement is the maximum possible.

The state (\ref{s0110}) possesses one ebit of bipartite entanglement not only between any two pairs of modes (chosen from $ABCD$) but between any one single mode and the remaining three. As a curiosity we note that under unitary redefinitions the minimum bipartite entanglement between one mode and three others is no longer 1 ebit but less, $E_{{\rm min}}=2-3\log_2(3)/4\approx 0.8113$. In this minimum, the reduced density matrix of the single mode is
\begin{eqnarray}
\rho=\frac{1}{4}[
3|0\rangle\langle 0|+
|2\rangle\langle 2|].
\end{eqnarray}
The maximum possible entanglement is  $E_{{\rm max}}\approx 1.3002$, which was found numerically.

\subsection{$|100001\rangle+|010010\rangle+|001100\rangle$}\label{orb}
Continuing the series with two photons in 6 modes where the state has the symmetric form
\begin{eqnarray}\label{s001100}
|\Psi\rangle=(|0\rangle|0\rangle|1\rangle|1\rangle|0\rangle|0\rangle
+|0\rangle|1\rangle|0\rangle|0\rangle|1\rangle|0\rangle\nonumber\\
+|1\rangle|0\rangle|0\rangle|0\rangle|0\rangle|1\rangle
)/\sqrt{3},
\end{eqnarray}
generalizing (\ref{s0110}),
we might conclude from writing the state as
\begin{equation}\label{s001100c}
|\Psi\rangle=(|A\rangle|F\rangle+|B\rangle|E\rangle+|C\rangle|D\rangle)
/\sqrt{3},
\end{equation}
that we have $\log_2(3)$ ebits of entanglement, since each photon ``has a three-dimensional Hilbert space attached to it'', to use the incorrect picture once again. 

However, restricting ourselves to bipartite entanglement between two triplets of modes, we have $E_{{\rm min}}=1$ and $E_{{\rm max}}=\log_2(5)$. This is the direct generalization of the result of the preceding subsection. Here, $\log_2(5)$ ebits is the maximum possible because the reduced density matrix $\rho_{A',B',C'}$ can be at most of rank 5. 

We in fact conjecture that for states with 2 photons in $2N$ modes of the form (\ref{s001100c}), or (\ref{s001100}), (which appear to have $\log_2(N)$ ebits of entanglement) we always have $E_{{\rm min}}=1$ and $E_{{\rm max}}=\log_2(N+2)$, if we consider only entanglement between two sets of $N$ modes. Note that the reduced density matrix of $N$ modes in a state with exactly $N$ photons can be at most of rank $N+2$: there are $N$ different ways in which 1 photon can be distributed over $N$ modes, and there are 2 ways either set of $N$ modes can have no photons. The case $N=8$ was considered in \cite{double}, where the entanglement was called ``double entanglement'', referring to the fact that the entanglement appears to be equal to 2 ebits.

Let us conclude this subsection by discussing briefly physical implementations of a state like (\ref{s001100}).
The six modes $A\ldots F$ could, of course, correspond to different spatial modes, or to two orthogonally polarized sets of three different spatial modes. But a recent experiment \cite{zeil} suggests one may alternatively make use of a different degree of freedom, namely the transverse mode profile: Within the paraxial approximation, a beam propagating in a given ($z$) direction, with fixed polarization and fixed frequency, still has an infinity of degrees of freedom left. For instance, one complete set of paraxial mode functions is given by the Hermite-Gaussian (HG) modes, another by the Laguerre-Gaussian (LG) modes \cite{siegman}. The latter modes may be characterized by a quantum number $L_z$, describing orbital angular momentum of light \cite{allen}. It is not straightforward to implement arbitrary unitary operations of the form (\ref{redef}) in this case. However, specific transformations are known: the conversion of HG modes into LG modes of the same order and {\em vice versa} with astigmatic lenses was discussed in Refs.~\cite{enkbey}, and a larger set of unitary operations is considered in Ref.~\cite{allenb}. Entangled states using the transverse degrees of freedom were discussed in \cite{downorb}.

\subsection{$|0220\rangle+|2002\rangle-|1111\rangle$}
A 4-photon state of the form 
\begin{equation}\label{s0220}
|\Psi\rangle=(|0\rangle|2\rangle|2\rangle|0\rangle
+|2\rangle|0\rangle|0\rangle|2\rangle- 
|1\rangle|1\rangle|1\rangle|1\rangle
)/\sqrt{3},
\end{equation}
features in a recent experiment \cite{dik}. There are two preferred pairs of modes in the experiment (the first two and the last two) and so it is correct to infer that the state (\ref{s0220}) possesses $\log_2(3)$ ebits of entanglement between those two pairs of modes. Restricting ourselves to bipartite entanglement between pairs of modes, this is in fact equal to the minimum, $E_{{\rm min}}=\log_2(3)$. By numerically searching for the maximum we found that $E_{{\rm max}}\approx 2.9798$, which is obtained when the reduced density matrix of a pair of modes is of the largest possible rank 9 \cite{Schmidt4}, but not maximally mixed ($\log_2(9)\approx 3.17$). 

Since in the experiment of Ref.~\cite{dik} the 4 modes involved are orthogonally polarized signal and idler modes, only ordinary and polarizing beam splitters are needed to transform the original state with $\log_2(3)\approx 1.585$ ebits into one with almost 3 (in fact, $E_{{\rm max}}$) ebits of entanglement. 
\subsection{$|00\rangle+|11\rangle$}
Our final example is a superposition of states with different photon numbers,
\begin{equation}\label{s0011}
|\Psi\rangle=(|0\rangle|0\rangle+|1\rangle|1\rangle)/\/\sqrt{2}.
\end{equation}
Again, this example shows that the entanglement is not between photons since the presence of the vacuum is here an essential part of the entanglement.
Written in the form (\ref{s0011}) we would conclude we have 1 ebit of entanglement. 
However, for the minimum amount of entanglement we find $E_{{\rm min}}\approx 0.3546$. More precisely, we get $E_{{\rm min}}=-\sum_k \lambda_k\log_2\lambda_k$ with 
$\lambda_{1,2}=1/2\pm\sqrt{3}/4$ the eigenvalues of the reduced density matrix of a single mode if we rewrite the state as 
\begin{equation}\label{s0011b}
|\Psi\rangle=(|0\rangle|0\rangle+(|0\rangle|2\rangle+|2\rangle|0\rangle)/\sqrt{2})/\sqrt{2}, 
\end{equation}
which is obtained by the same transformation (\ref{bishop}) as in Subsection \ref{pawn}.
The maximum amount of entanglement is, surprisingly perhaps, just a tiny little bit over 1 ebit, namely $E_{{\rm max}}\approx 1.0071$, as was found numerically.
\section{Conclusions}
There are two differences between a second-quantized theory, such as Quantum Electrodynamics, and a first-quantized theory, such as standard non-relativistic quantum mechanics of material particles. One is that particles, photons in particular, are not systems in a second-quantized theory. The other difference relevant to quantum-information processing is that the systems, the EM field modes in particular, can be redefined arbitrarily, at least in principle. But one does not have any freedom in (re)defining ions in an ion trap or atoms inside a cavity. (Of course, in a second-quantized theory of matter particles this would be different, but we live in a world where non-relativistic quantum theory is an extremely good and convenient approximation for atoms and ions.)

It turns out that the amount of entanglement present in a given state depends on how one defines one's systems. 
The unitary transformations (\ref{redef}) may be viewed as, indeed, merely redefinitions of field modes. In practice, however, there is a more useful point of view. The transformation (\ref{redef}) may be seen as a unitary operation on modes that can actually be performed by linear optics elements such as beam splitters, polarizing beam splitters, or in more complicated cases by astigmatic lenses. This way the entanglement in a given state may often be increased beyond the amount that is ``obviously'' present in the state, as was shown in several examples.

Finally, let us point out some relations with previous work. The conclusions reached in the present paper agree, in some indirect sense, with those in two quite different papers. First, in Ref.~\cite{spreeuw} it is shown that certain aspects of {\em local} entanglement, that is, entanglement between different modes that are all in the same location, can be simulated with classical light beams. 
Second, in Ref.~\cite{scudo} it is shown that the amount of entanglement between the spatial and spin degrees of freedom of a relativistic spin-1/2 particle depends on the reference frame of the observer.
\section*{Acknowledgments}It's a pleasure to thank Chris Fuchs, Gerard Nienhuis, and Rob Pike for useful comments.


\begin{thebibliography}{99}
\bibitem{E}The mode functions and mode operators appear together in, for example, the expression for the electric field operator in the Heisenberg picture, 
\[
\vec{E}=\sum_j i{\cal E}_j \big[\vec{F}_{j}(\vec{r}) a_j(t) -
\vec{F}^*_{j}(\vec{r}) a^{\dagger}_j(t) \big],
\]
with ${\cal E}_j=(\hbar\omega_j/2\epsilon_0)^{1/2}$ a normalization constant and $\omega_j$ the frequency of the mode. For convenience we used a notation for discrete quantum numbers $j$ but some of them may actually be continuous. The mode functions further satisfy (because of the Maxwell equations)
\[\nabla^2\vec{F}_j=-\left(\frac{\omega_j}{c}\right)^2\vec{F}_j
\]
and the transversality condition
\[\vec{\nabla}\cdot\vec{F}_j=0.\]
\bibitem{E2}and so that physical quantities, such as the electric field $\vec{E}$, are invariant.
\bibitem{Schmidt4}There are at most 9 terms in the Schmidt decomposition: the 4 photons can be distributed among the two pairs of modes in 9 different ways. There is 1 state with a distribution (0,4), 2 with (1,3), 3 with (2,2), 1 with (3,1), and 1 with (4,0).
\end{thebibliography}
\end{document}